\documentclass[aps,preprintnumbers,eqsecnum,amsmath,amssymb,showpacs,nofootinbib]{revtex4}
\usepackage{graphicx}\usepackage{dcolumn}\usepackage{bm}\usepackage{epsfig}

\begin{document}

\title{Strong three-meson couplings of $J/\psi$ and $\eta_c$}
\author{Wolfgang Lucha$^a$, Dmitri Melikhov$^{b,c}$, Hagop Sazdjian$^d$, and Silvano Simula$^e$}
\affiliation{$^a$Institute for High Energy Physics, Austrian Academy of Sciences, Nikolsdorfergasse 18, A-1050 Vienna, Austria\\
$^b$D.~V.~Skobeltsyn Institute of Nuclear Physics, M.~V.~Lomonosov
Moscow State University, 119991, Moscow, Russia\\ 
$^c$Faculty of Physics, University of Vienna, Boltzmanngasse 5, A-1090 Vienna, Austria\\
$^d$IPN, CNRS/IN2P3, Universit\`e Paris-Sud 11, F-91406 Orsay, France\\ 
$^e$INFN, Sezione di Roma III, Via della Vasca Navale 84, I-00146 Roma, Italy }
\date{\today}

\begin{abstract}
We discuss the strong couplings $g_{PPV}$ and
$g_{VVP}$ for vector ($V$) and pseudoscalar ($P$) mesons, at least
one of which is a charmonium state $J/\psi$ or $\eta_c$. The
strong couplings are obtained as residues at the poles of suitable
form factors, calculated in a broad range of momentum transfers using a dispersion formulation of 
the relativistic constituent quark model. The form factors obtained in this approach 
satisfy all constraints known for these quantities in the heavy-quark limit. 
Our results suggest sizably higher values for the strong meson couplings than those reported in the 
literature from QCD sum rules.
\end{abstract}

\maketitle

\section{Introduction}
Strong couplings involving three mesons are
complicated objects posing a great challenge for their theoretical
study. The $D^*D\pi$ coupling, for which most theoretical analyses
predicted values sizably smaller than the one later measured by
CLEO \cite{cleo}, illustrates this statement very well. In this
letter, we address the strong three-meson couplings involving
$J/\psi$ and $\eta_c$ states. These quantities cannot be measured
directly in strong $J/\psi$ and $\eta_c$ decays, but they are
important for our understanding of the $J/\psi$ and $\eta_c$
properties in a hadronic medium \cite{lin2000}.

Most results for charmonium couplings arose from rather detailed
QCD sum-rule calculations
\cite{bracco2005,matheus2005,bracco2014,bracco2015}. In the past,
however, the application of QCD sum rules to three-meson couplings
faced a great problem: QCD sum rules strongly underestimated the
$D^*D\pi$ coupling (see, e.g., \cite{braun}) and the origin of
this discrepancy has not been fully clarified. We thus present an
alternative analysis of the family of $J/\psi$ and $\eta_c$
couplings using the relativistic dispersion approach \cite{da},
one of the approaches which managed to predict correctly the
$D^*D\pi$ coupling \cite{mb,ms} before the CLEO measurement.

The strong couplings in the focus of our interest, $g_{PV'V}$ and
$g_{PP'V},$ are defined by
\begin{align}
\label{g}\langle P'(p_2)V(q)|P(p_1)\rangle&=
-\mbox{$\frac{1}{2}$}g_{PP'V}(p_1+p_2)^\mu\varepsilon^*_\mu(q),\nonumber\\
\langle V'(p_2)V(q)|P(p_1)\rangle&
=-\epsilon_{\varepsilon^*(q)\varepsilon^*(p_2) p_1 p_2}g_{PV'V},
\end{align}
with momentum transfer $q=p_1-p_2$. Accordingly, $g_{PP'V}$ is
dimensionless whereas $g_{PV'V}$ has inverse mass dimension.

These strong couplings are related to the residues of the poles in
the transition form factors at time-like momentum transfer arising
from contributions of intermediate meson states in the transition
amplitudes' $q^2$ channel. We study the form factors $F_+^{P\to
P'}(q^2)$, $V^{P\to V}(q^2)$, and $A_0^{P\to V}(q^2),$ related to
the transition amplitudes induced by vector quark currents $\bar
q_2\gamma_\mu q_1$ or axial-vector quark currents $\bar
q_2\gamma_\mu\gamma_5 q_1$:
\begin{align*}
\langle P'(p_2)|\bar q_2\gamma_\mu q_1|P(p_1)\rangle&=F_+^{P\to
P'}(q^2)(p_1+p_2)_\mu+\cdots,\\\langle V(p_2)|\bar q_2\gamma_\mu
q_1|P(p_1)\rangle&=\frac{2V^{P\to
V}(q^2)}{M_P+M_V}\epsilon_{\mu\varepsilon^*(p_2)p_1 p_2},\\\langle
V(p_2)|\bar q_2\gamma_\mu\gamma_5q_1|P(p_1)\rangle&={\rm
i}q_\mu(\varepsilon^*(p_2)p_1)\frac{2M_V}{q^2}A_0^{P\to
V}(q^2)+\cdots, 
\end{align*}
where dots stand for other Lorentz structures. The poles in the above form factors are of the form
\begin{eqnarray}
\label{residues}
F_+^{P\to P'}(q^2)&=&\frac{g_{PP'V_R}f_{V_R}}{2M_{V_R}}\frac{1}{1-q^2/M_{V_R}^2}+\cdots,\nonumber\\
V^{P\to V}(q^2)&=&\frac{(M_V+M_P)g_{PVV_R}f_{V_R}}{2M_{V_R}}\frac{1}{1-q^2/M^2_{V_R}}+\cdots, \nonumber\\
A_0^{P\to V}(q^2)&=&\frac{g_{PP_RV}f_{P_R}}{2M_{V}}\frac{1}{1-q^2/M^2_{P_R}}+\cdots.
\end{eqnarray}
In these relations, $P_R$ and $V_R$ label pseudoscalar and vector
resonances with appropriate quantum numbers; $f_P$ and~$f_V$ are
the leptonic decay constants of the pseudoscalar and vector
mesons, respectively, defined in terms of the amplitude of the
meson-to-vacuum transition induced by the axial-vector or vector
quark currents according to
\begin{align*}
\langle0|\bar q_1\gamma_\mu\gamma_5q_2|P(p)\rangle&={\rm
i}f_Pp_\mu,\\\langle0|\bar q_1\gamma_\mu
q_2|V(p)\rangle&=f_VM_V\varepsilon_\mu(p).
\end{align*}

\section{Dispersion formulation of the relativistic constituent quark model}
Relativistic constituent quark models \cite{cqm}
proved to constitute an efficient tool for the study of hadron
properties,~in particular of meson decay constants and transition
form factors. An essential feature of the constituent quark
picture is the appropriate matching of the quark currents in QCD
($\bar q\gamma_\mu q$, $\bar q\gamma_\mu\gamma_5 q$, etc.) and the
associated currents formulated in terms of constituent quarks
($\bar Q\gamma_\mu Q$, $\bar Q\gamma_\mu\gamma_5 Q$, etc.). For
light quarks, for instance, partial conservation of the
axial-vector current requires the appearance of the pseudoscalar
structure in the axial-vector current of the constituent quarks,
similar to the case of the axial-vector current of the nucleon
\cite{gv}. For the currents containing heavy quarks,~the matching
conditions are simpler:
\begin{align*}
\bar q_1\gamma_\mu q_2&=g_V\bar Q_1\gamma_\mu Q_2+\cdots,\\\bar
q_1\gamma_\mu\gamma_5q_2&=g_A\bar Q_1\gamma_\mu\gamma_5Q_2+\cdots,
\end{align*}
where the dots indicate contributions of other possible Lorentz
structures \cite{gv}. Constituent quarks $Q_1$ and $Q_2$ have masses $m_1$ and $m_2$, respectively. 
In general, the form factors $g_V$ and $g_A$ depend on the momentum transfer. Vector current conservation
requires $g_V=1$ at zero momentum transfer for the elastic current
and at zero recoil for the heavy-to-heavy quark transition. The
specific values of the form factors $g_V$ and $g_A$ and their
momentum dependences belong to the parameters of the model, as well as the quark masses and the wave functions 
of mesons regarded as relativistic quark--antiquark bound states. 
A relativistic treatment of two-particle contributions to the
bound-state structure may be consistently formulated within a
relativistic dispersion approach which takes into account only
two-particle intermediate quark-antiquark states in Feynman diagrams \cite{anis}.
Such a formulation is explicitly relativistic-invariant: hadron observables like form factors or decay constants 
are given by spectral representations over the invariant masses of the quark-antiquark intermediate states. 
Application of the dispersion formulation of the constituent quark picture to heavy-to-light meson form factors has
convincingly demonstrated the reliability of this approach \cite{ms}.

\subsection{Meson decay constants and form factors as spectral integrals}
Within the dispersion formulation of the constituent quark model, the decay constants
$f_P$ and $f_V$ of pseudoscalar and vector mesons are expressed~in
the form of relativistic spectral representations, over the
invariant masses of the intermediate quark--antiquark states,~of the spectral densities involving the nonperturbative 
meson wave functions $\phi_P(s)$ and $\phi_V(s),$ respectively \cite{da}:
\begin{eqnarray}
\label{fpv}
f_P&=&\sqrt{N_c}\int\limits_{(m_1+m)^2}^\infty
{\rm d}s\,\phi_P(s)\,(m_1+m)\frac{\lambda^{1/2}(s,m_1^2,m^2)}{8\pi^2s}\frac{s-(m_1-m)^2}{s}, \nonumber
\\
f_V&=&\sqrt{N_c}\int\limits_{(m_1+m)^2}^\infty
{\rm d}s\,\phi_V(s)\,\frac{2\sqrt{s}+m_1+m}{3}\frac{\lambda^{1/2}(s,m_1^2,m^2)}{8\pi^2s}\frac{s-(m_1-m)^2}{s},
\end{eqnarray}
with $\lambda(a,b,c)\equiv(a+b-c)^2-4ab$. The wave functions
$\phi_i(s)$, $i=P,V,$ can be written as
\begin{eqnarray}
\label{k2}
\phi_i(s)=\frac{\pi}{\sqrt2}
\frac{\sqrt{s^2-(m_1^2-m^2)^2}}{\sqrt{s-(m_1-m)^2}}
\frac{w_i(k^2)}{s^{3/4}},\qquad
k^2=\frac{\lambda(s,m_1^2,m^2)}{4s},
\end{eqnarray}
with $w_i(k^2)$ normalized according to
\begin{eqnarray}
\label{norma}
\int{\rm d}k\,k^2w^2_i(k^2)=1.
\end{eqnarray}
Notice that Eqs.~(\ref{fpv}) may be rewritten as the Fourier transform of the meson relativistic wave function at the origin. 

\begin{figure}[!t]
\begin{tabular}{ccc}
\includegraphics[width=4.cm]{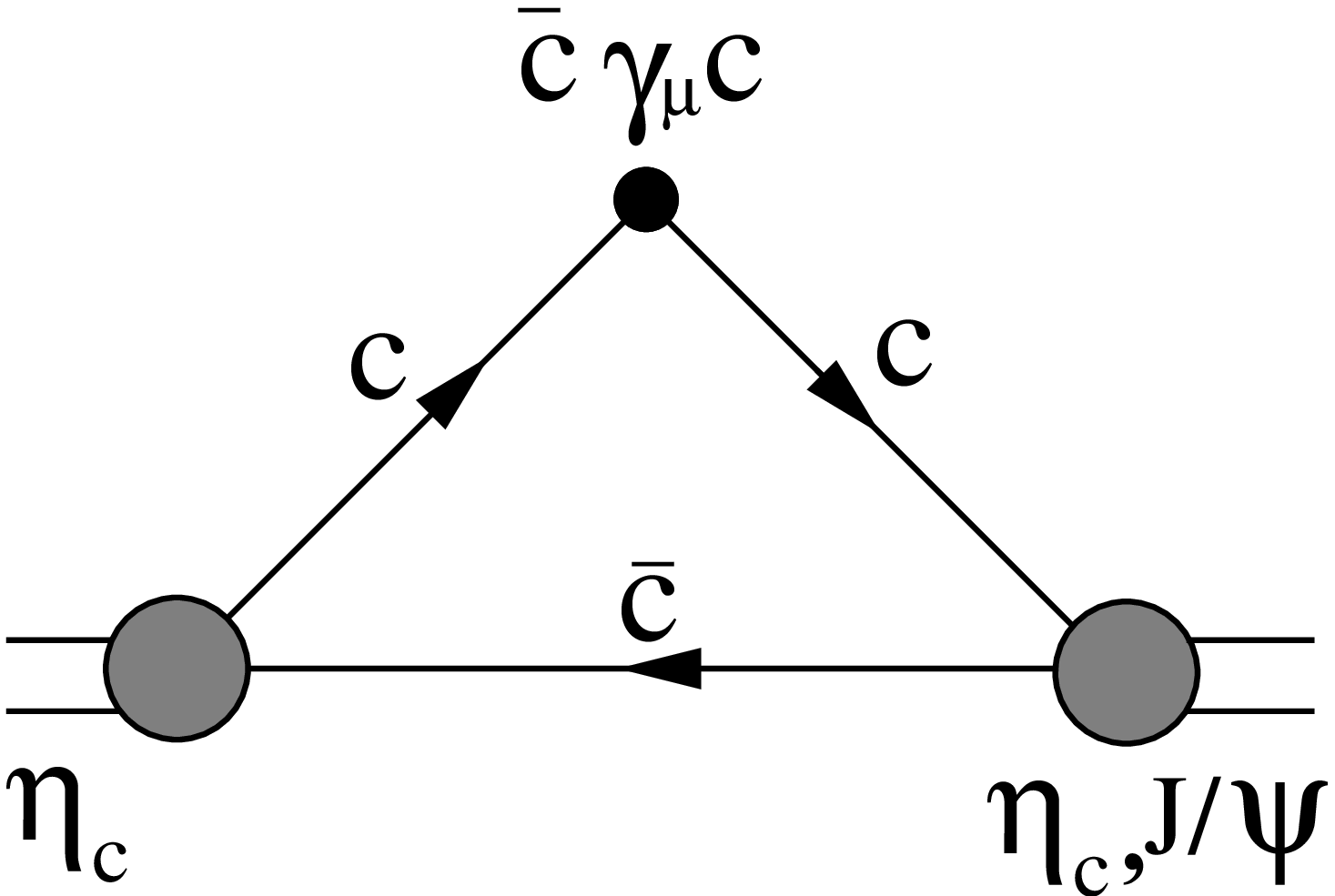}$\quad$&$\quad$
\includegraphics[width=4.cm]{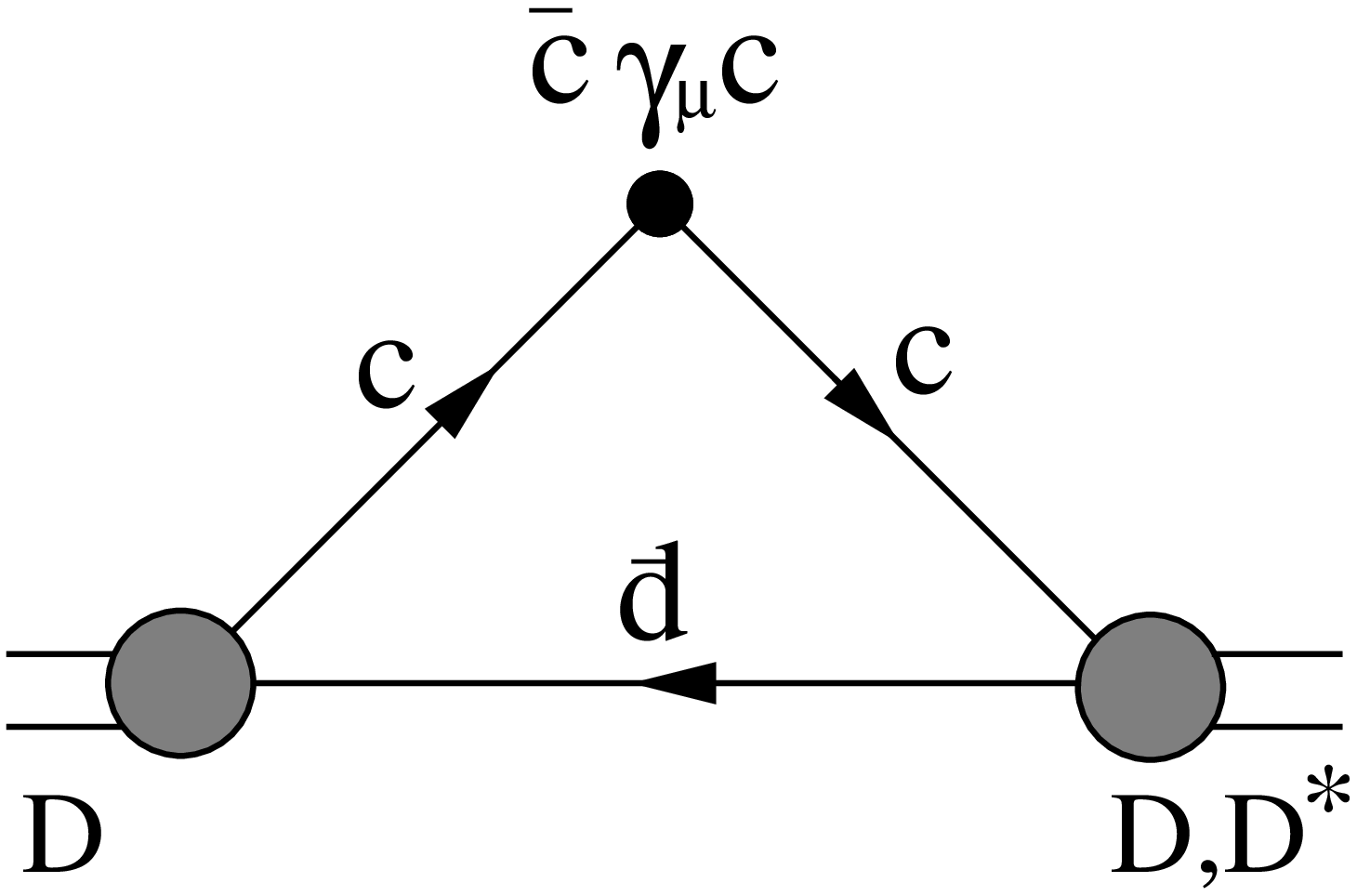}$\quad$&$\quad$
\includegraphics[width=4.cm]{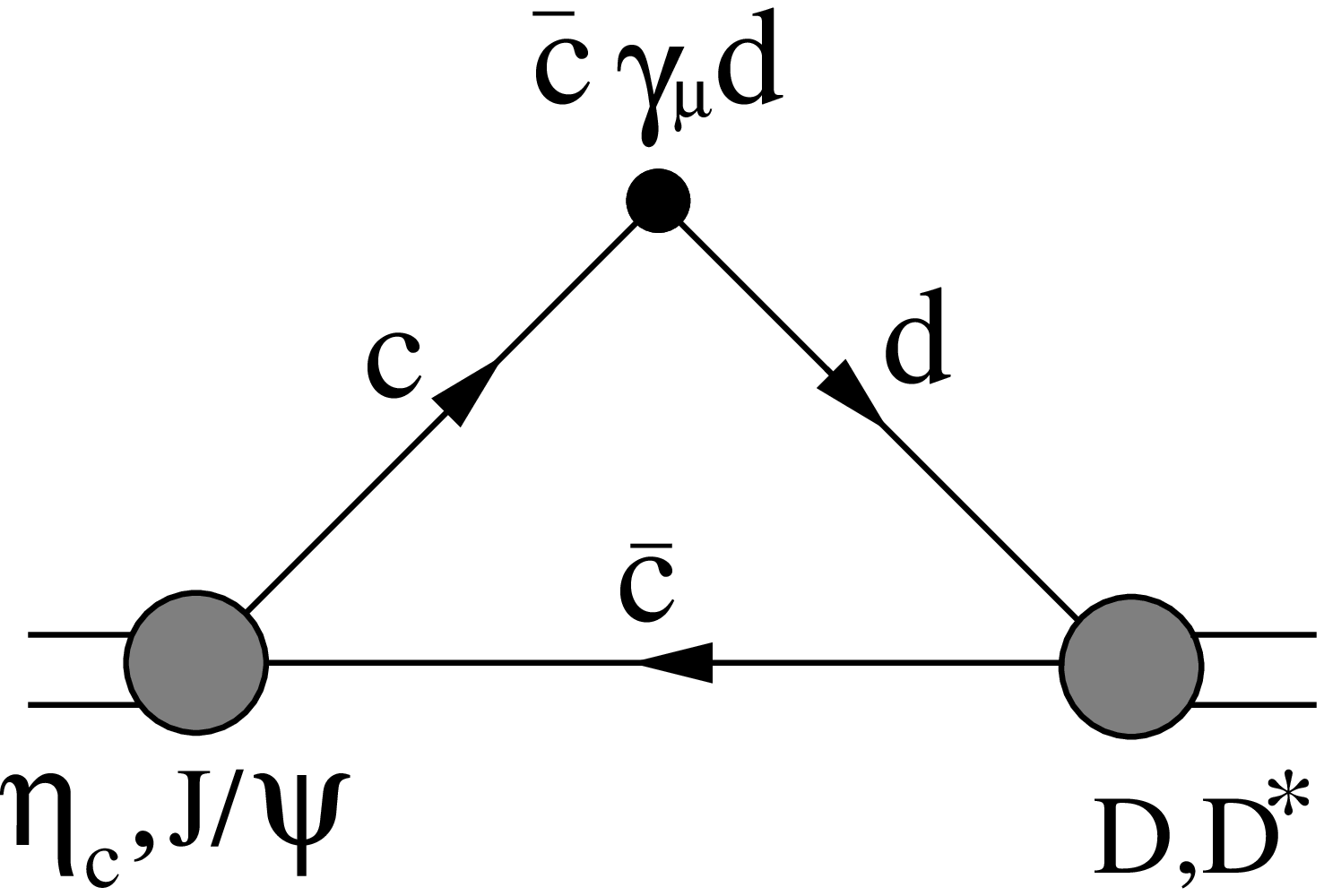}\\
(a)$\quad$&$\quad$(b)$\quad$&$\quad$(c)\end{tabular}
\caption{\label{Plot:1} Feynman diagrams for the transitions under
consideration induced by the quark vector currents $\bar
Q\gamma_\mu Q$: (a) $\eta_c\to\eta_c,J/\psi$ induced by the
current $\bar c\gamma_\mu c,$ (b) $D\to D,D^*$ induced by the
current $\bar c\gamma_\mu c,$ and (c) $\eta_c,J/\psi\to D,D^*$
induced by the current~$\bar c\gamma_\mu d.$}
\end{figure}

%******************************************
Similarly, the $M_1(p_1)\to M_2(p_2)$ transition form factors
induced by the constituent-quark transition current $\bar Q_1\hat OQ_2$~in the kinematical region $-\infty<q^2\le(m_2-m_1)^2$ is
given by the double spectral representation 
\begin{equation}
\label{ff}
F_i(q^2)=\int{\rm d}s_1\,\phi_1(s_1)\int{\rm d}s_2\,\phi_2(s_2)\,\Delta_i(s_1,s_2,q^2).
\end{equation}
The function $\Delta_i(s_1,s_2,q^2)$ is the double spectral density of the relevant Feynman diagrams with constituent~quarks
in the loop (Fig.~\ref{Plot:1}). It contains the $\theta$-functions corresponding to the quark-antiquark thresholds and 
a specific constraint coming form the triangle Feynman diagram. The explicit expressions for
$\Delta_i(s_1,s_2,q^2)$ are given in Sect.~3.2 of \cite{da} and will not be reproduced here. 
We point out that at $q^2<0$, the form factors obtained within the dispersion formulation are equal to the 
form factors of the light-front relativistic constituent quark model (Ref.~3 in \cite{cqm}). 
Correspondingly, the double spectral representation (\ref{ff}) at $q^2<0$ may be rewritten as the convolution of the light-cone wave 
functions of the initial and the final hadrons (see Eq.~(2.86) of \cite{da}), or, equivalently, as the Fourier transform in the 
transverse variables of the overlap of these wave functions. The merit of having explicitly 
relativistic-invariant spectral representations compared to other formulations is the possibility to obtain the form factors 
in the decay region $0<q^2\le(m_2-m_1)^2$ by the analytic continuation in $q^2$ which was shown to lead to the appearance of the 
anomalous cut \cite{melikhov}. Noteworthy, both the normal and the anomalous contributions involve the $s_1$ and $s_2$ integrations 
over the corresponding two-particle cuts, i.e. for $k_1^2>0$ and $k_2^2>0$ with $k_{_1,2}$ given by (\ref{k2}). 
As the result, the form factors in a broad kinematical region $-\infty<q^2\le(m_2-m_1)^2$ are 
expressed in terms of the relativistic wave functions of the participating mesons $w_1(k_1^2)$ and $w_2(k_2^2)$ 
with $k_1^2>0$ and $k_2^2>0$. In this region, the precise form of the wave functions $w(k_i^2)$ 
is not crucial; essential is only that the confinement effects have been taken into account. That is why, as shown 
in the applications to meson transition form factors \cite{ms}, a simple Gaussian parameterization can be adopted: 
\begin{equation}
\label{gauss} 
w_i(k^2)\propto\exp(-k^2/2\beta_i^2).
\end{equation}
The spectral representation (\ref{ff}) is based on constituent-quark degrees of freedom and we apply it to calculate the 
form factors in the region $q^2<(m_2-m_1)^2$. We then numerically interpolate the results of our calculations and use the 
obtained parameterizations to study the form factors at $q^2>(m_1-m_2)^2$, where one expects the appearance of meson resonance  
at $q^2=M_R^2$.  

The use of the dispersion formulation of the constituent quark model allows us to reveal the intimate connection between
different decay modes and to perform the calculations in the broad range of $q^2$ which includes the 
scattering region $q^2<0$ and the physical region of the quark weak decay $0<q^2<(m_2-m_1)^2$. In fact, 
quark models are the only approach that leads to relations between the decays of different mesons
through the meson wave functions~and provides the form factors in
the $q^2$-range indicated above. It is important to emphasize that the form factors (\ref{ff})
reproduce correctly the structure of the heavy-quark expansion in
QCD for heavy-to-heavy and heavy-to-light transitions if the
radial wave functions $w_i(k^2)$ are localized in a region of the
order of the confinement scale $\Lambda,$ i.e., $k^2\le\Lambda^2$ \cite{melikhov}. 

%********************************************************************************

\subsection{Parameters of the model}
For the wave functions, we make use of the simple Gaussian wave-function Ansatz which satisfies the localization requirement 
for $\beta\simeq \Lambda_{\rm QCD}$ and proved to provide a reliable picture of a
large class of transition form factors \cite{ms}.

Noteworthy, the quark-model double spectral representations take into account long-range QCD effects but not the short-range
perturbative corrections. However, the parameters of the model
(quark masses and nonperturbative meson wave functions
corresponding to the choice of the constituent-quark couplings
$g_V=1$ and $g_A=1$) are~assumed such that our dispersion approach
reproduces the observables (decay constants and some
``well-measured'' form factors~from lattice QCD); therefore,
radiative corrections to the quark propagators and to the vertices
at the moderate momentum transfers considered are effectively
taken care of by the use of constituent quark
masses\footnote{Indications of the appearance of the effective
constituent quark masses in the soft region come from several
different approaches \cite{constituent}.} and the meson wave
functions.

We employ the same values of the constituent quark masses and the
quark couplings that have been obtained in~\cite{ms}:
\begin{equation}
\label{quark_masses}g_V=g_A=1,\qquad m_d=m_u=0.23\;{\rm GeV},
\qquad m_s=0.35\;{\rm GeV},\qquad m_c=1.45\;{\rm GeV}.
\end{equation}
With the above quark couplings and masses, and the meson
wave-function parameters $\beta$ collected in
Table~\ref{table:parameters}, the~decay constants from our
dispersion approach reproduce the best-known decay constants of
pseudoscalar and vector mesons, also summarized in
Table~\ref{table:parameters}.

\begin{table}[t]
\caption{\label{table:parameters}Masses \cite{pdg},
leptonic decay constants and corresponding wave-function
parameters $\beta$ of charmed mesons and charmonia.}\centering
\begin{tabular}{|c|c|c|c|c|c|c|}\hline
& $D$ & $D^*$ & $D_s$ & $D_s^*$ & $\eta_c$ & $J/\psi$ \\\hline $M$
(GeV) & $1.87$ & $2.010$ & $1.97$ & $2.11$ & $2.980$ & $3.097$\\
$f$ (MeV) & $206\pm8$ \cite{lmsfD,pdg} & $260\pm10$
\cite{lmsfD*,fDDs*lat} & $248\pm2.5$ \cite{fetac_lat} & $311\pm9$
\cite{fpsi_lat} & $394.7\pm2.4$ \cite{fetac_lat} & $405\pm7$
\cite{fpsi_lat,pdg}\\ \hline $\beta$ (GeV) & $0.475$ & $0.48$ &
$0.545$ & $0.54$ & $0.77$ & $0.68$\\\hline\end{tabular}\end{table}

Using the parameter values (\ref{quark_masses}) and Table
\ref{table:parameters}, the spectral representations (\ref{ff})
yield the form factors numerically. We then interpolate our numerical results 
by a simple physically motivated formula
\begin{equation}
\label{fit1}
F(q^2)=\frac{F(0)}{\left(1-q^2/M_R^2\right)
\left(1-\sigma_1 q^2/M_R^2+\sigma_2q^4/M_R^4\right)},
\end{equation}
where $M_R=M_V$ for $F_+$ and $V,$ and $M_R=M_P$ for $A_0$. 
We may use the parameters $F(0)$, $\sigma_{1,2}$ and $M_R$ as parameters of our fitting procedures. It turns out that for 
all form factors considered in this work, the value of $M_R$ obtained by the fit turns out to be very close 
(within few percent accuracy) to the mass of the 
resonance with the appropriate quantum numbers. This property opens the possibility to use the obtained parameterization 
(\ref{fit1}) up to $q^2=M_R^2$ and to estimate the pole residues. In what follows we set $M_R$ equal to the known mass of 
the physical resonance and use the remaining parameters $F(0)$ and $\sigma_{1,2}$ as parameters of the fit. 
The parameters in (\ref{fit1}) are related to the pole residue via 
\begin{eqnarray}
{\rm Res}\,F(q^2=M_R^2)=\frac{F(0)}{1-\sigma_1+\sigma_2}. 
\end{eqnarray} 
The residue is given by products of the (known) weak and the strong couplings $g$ to be
determined. Finally, our fit parameters are $F(0)$, $\sigma_1,$ and the
strong coupling $g$ related to ${\rm Res}\, F(q^2=M_R^2)$.
In some cases, the residues of~different form factors involve the same strong coupling; for such form-factor sets  
a constrained interpolation will be done.

%*****************************************************
\section{The $\eta_c\eta_cJ/\psi$ and $\eta_cJ/\psi J/\psi$ strong couplings}
The double spectral representations enable us to
calculate the necessary form factors as soon as the vertex
functions of the $\eta_c$ and $J/\psi$ are given.
%The $c$-quark constituent mass has been determined in \cite{ms}.
We fix the wave-function slope parameters $\beta_i$ such~that the
decay constants of $\eta_c$ and $J/\psi$ are reproduced by the
spectral representations (\ref{fpv}). Using for $\eta_c$ the
lattice finding $f_{\eta_c}=(394.7\pm2.4)\;\mbox{MeV}$
\cite{fetac_lat} and for $J/\psi$ the experimental result
$f_\psi=(407\pm5)\;\mbox{MeV}$ \cite{pdg} --- which agrees
excellently with the lattice determination
$f_\psi=(405\pm6\pm2)\;\mbox{MeV}$ \cite{fpsi_lat} --- yields the
wave-function parameters $\beta_{\eta_c}=0.77$ GeV and
$\beta_\psi=0.68$ GeV. As soon as these are fixed, we calculate
the form factors $F_+(\eta_c\to\eta_c),$ $V(\eta_c\to\psi),$ and
$A_0(\eta_c\to\psi)$~in the kinematical region $q^2<0$ by using
the dispersion representations (\ref{ff}).

The $\eta_c$ elastic form factor is normalized to
$F^{\eta_c\to\eta_c}_+(0)=1$ by elastic vector-current
conservation. Our determination of the $\psi\to\eta_c$ transition
form factor $V^{\eta_c\to\psi}(0)=1.80$, describing the
$\psi\to\eta_c\gamma$ radiative transition, is in reasonable
agreement with both the data \cite{V(0)exp} and the lattice-QCD
result \cite{V(0)lat}, in spite of some tension between these two
findings: $V^{\rm exp}(0)=1.68\pm0.14$ vs.\ $V^{\rm
lat}(0)=1.92\pm0.03\pm0.02$. N.B.: In the limit $m_c\to\infty$, the heavy-quarkonium transition form factor approaches the value
$V(0)=2$.

Next, we interpolate the results of our form-factor calculations performed for $-M^2_{\psi}<q^2<0$, by the fit formula~(\ref{fit1}).
The residues of the form factors $F_+(\eta_c\to\eta_c)$ and $A_0(\eta_c\to\psi)$ are given in terms of one and the same
coupling $g_{\eta_c\eta_c\psi}$:
\begin{eqnarray}
\nonumber
{\rm Res}\;F_+(q^2=M_\psi^2)
=g_{\eta_c\eta_c\psi}f_\psi/2M_\psi,\qquad{\rm
Res}\;A_0(q^2=M_{\eta_c}^2)=g_{\eta_c\eta_c\psi}f_{\eta_c}/2M_\psi.
\end{eqnarray}
Hence, we perform a combined fit to the two form factors
$F_+(\eta_c\to\eta_c)$ and $A_0(\eta_c\to\psi)$, regarding
$g_{\eta_c\eta_c\psi}$, $A_0^{\eta_c\to \psi}(0)$  
and the parameters $\sigma_1$ for $F_+(\eta_c\to\eta_c)$ and $A_0(\eta_c\to\psi)$ 
as the fit parameters (recall that $F_+^{\eta_c\to\eta_c}(0)=1$ due to current conservation). 
The corresponding results are given in Table \ref{table:c-loop}. These fits represent 
the numerical outcomes with a fantastic accuracy --- better than 0.2\% --- in the full $q^2$
range considered. This lends strong support to the reliability of~our approach to
charmonia, in spite of the approximate form of our wave-function model.

The excellent description~of~our calculated form factors by the interpolation formula (\ref{fit1}) suggests  
that this parameterization may be extended up to $q^2=M_R^2$ and used to calculate the 
strong couplings from the residue of the pole at $q^2=M_R^2$ in (\ref{fit1}).  

\begin{table}[!hb]
\caption{\label{table:c-loop}
Form factors describing the $\eta_c\to\eta_c$ and $\eta_c\to
J/\psi$ transitions and corresponding strong couplings. 
%Notice that the fit parameters $F(0)$, $\sigma_1$ and $g$ are rather strongly correlated with each other.
}\centering
\begin{tabular}{|c|cc|c|}\hline
Amplitude & $\langle\eta_c|\bar c\gamma_\mu c|\eta_c\rangle$ &
$\langle J/\psi|\bar c\gamma_\mu\gamma_5c|\eta_c\rangle$\ \ &
$\langle J/\psi|\bar c\gamma_\mu c|\eta_c\rangle$\\\hline Form
factor &\ $F_+(\eta_c\to\eta_c)$\ &\ $A_0(\eta_c\to J/\psi)$\ \ & $V(\eta_c\to J/\psi)$\\
\hline \vspace{-2.5ex}&&&\\ 
$F(0)$ & $1$ & $0.900\pm 0.004$ & $1.80\pm 0.01$\\ 
$M_R$ & $M_{\psi}$ & $M_{\eta_c}$ & $M_\psi$\\
$\sigma_1$ & $0.60\pm 0.01$ & $0.77\pm 0.02$ & $0.73\pm 0.04$\\ \hline\ Strong coupling\ \
& \multicolumn{2}{c|}{$g_{\eta_c\eta_c\psi}=25.8\pm1.7$} &\ $g_{\eta_c\psi\psi}=(10.6\pm1.5)\,{\rm
GeV}^{-1}$\ \
\\\hline
\end{tabular}
\end{table}
%Our estimate for the strong couplings of charmonia read 
%\begin{eqnarray}
%g_{\eta_c\eta_c\psi}=25.8\pm1.7,\qquad
%g_{\eta_c\psi\psi}=(10.6\pm1.5)\;{\rm GeV}^{-1}.
%\end{eqnarray}
The statistical uncertainty reflects merely the accuracy of the description of the calculation outcomes~by the fit
formula, but does not take into account the systematic uncertainties related to the approximate character of the model
and the specific form of the interpolating formula. The latter cannot be probed unambiguously. 
However, comparison with the results of the experiment or lattice QCD in those cases where these results are available, 
shows that the systematic uncertainty does not exceed the 10--15\% level. 

In the limit $m_c\to\infty$, we have $F_+(0)=1$ and
$V(0)=2.$ Consequently, the strong couplings of heavy quarkonia
satisfy $g_{\eta_c\eta_c\psi}=M_\psi g_{\eta_c\psi\psi},$ which is
fulfilled with 20\% accuracy for the charmonium couplings.

%************************************************************************
\section{Strong couplings of $\eta_c$ and $J/\psi$ to $D$ and $D^*$} 
Here, the couplings of interest may be extracted from the
residues of poles in form factors that describe two different
kinds of transitions: transitions between the charmed mesons,
induced by the currents $\bar c\gamma_\mu c$ and $\bar
c\gamma_\mu\gamma_5c$ (Fig.~\ref{Plot:1}b), and transitions
between the charmonia and the charmed mesons, induced by the
currents $\bar c\gamma_\mu d$ and $\bar c\gamma_\mu\gamma_5d$
(Fig.~\ref{Plot:1}c). Our results for the form factors and the
corresponding couplings are presented in Tables
\ref{table:ccd-loop1} and \ref{table:ccd-loop2}. 
Again, the small uncertainties of the obtained couplings do not reflect possible systematic errors related to 
the approximate nature of the dispersion approach.

\begin{figure}[t]
\includegraphics[width=8cm]{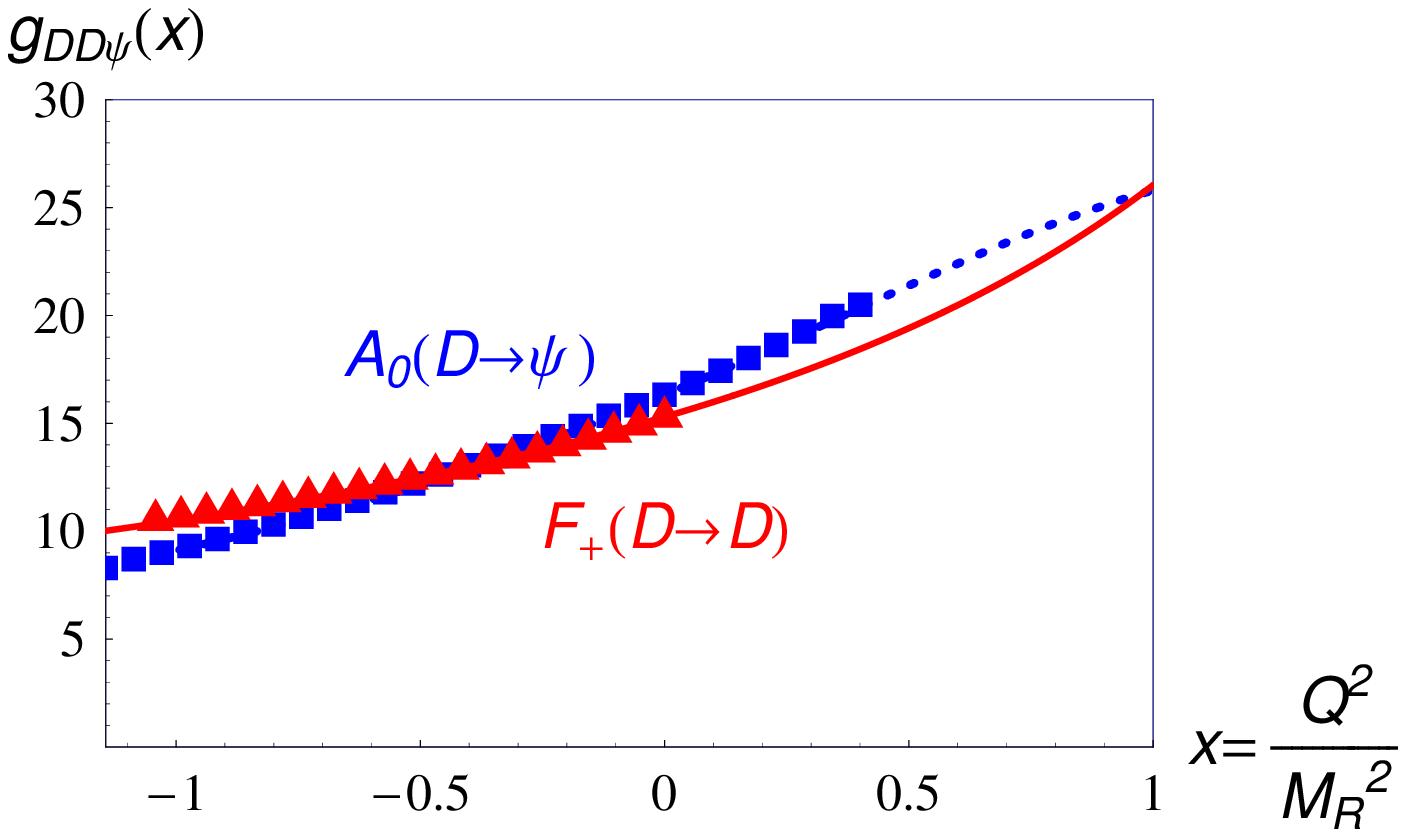}\includegraphics[width=8cm]{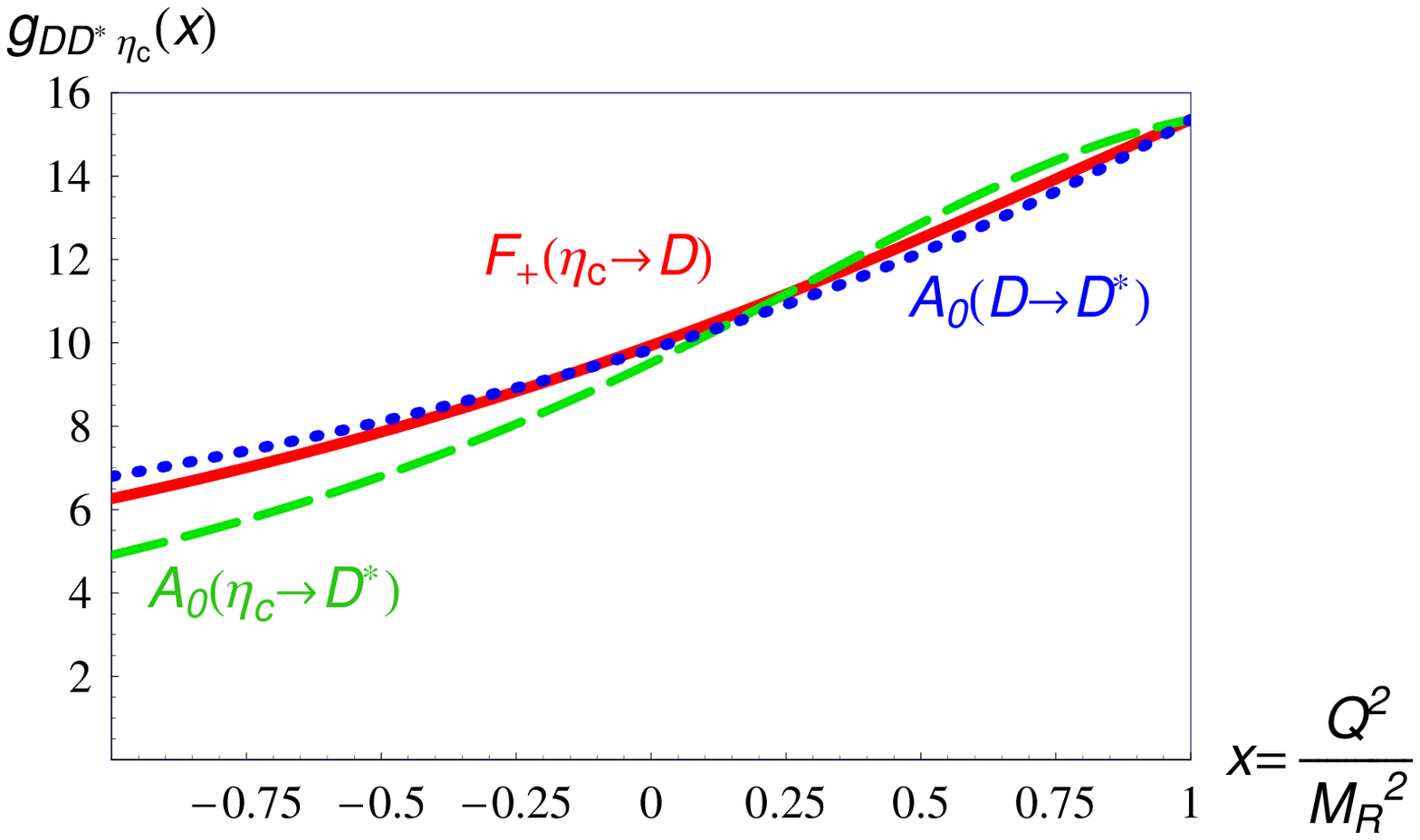}
\caption{\label{Plot:2}
The ``off-shell'' strong couplings. Left: 
$g_{D\hat{D}\psi}(x)=\frac{2M_\psi}{f_D}(1-x)A^{D\to\psi}_0(q^2)$,
$x=q^2/M_D^2$ (blue squares and blue dotted line),~and
$g_{DD\hat{\psi}}(x)=\frac{2M_\psi}{f_\psi}(1-x)F_+^{D\to D}(q^2)$, $x=q^2/M^2_\psi$ 
(red triangles and red solid line),
extracted from the form factors $A^{D\to\psi}_0(q^2)$~and
$F_+^{D\to D}(q^2)$, respectively. Triangles and squares indicate
the results computed numerically from the spectral
representations,~the dotted and solid lines the fits interpolating
the results and then used for extrapolation to the pole regions.
Right: 
$g_{D\hat{D^*}\psi}(x)$ obtained from $F_+^{\eta_c\to D}$ (red solid line); 
$g_{DD^*\hat{\psi}}(x)$ obtained from $A_0^{D\to D^*}$ (blue dotted line); 
$g_{\hat{D}D^*\psi}(x)$ obtained from $A_0^{D\to D^*}$ (green dashed line). }
\end{figure}

\begin{table}[!hbt]
\caption{\label{table:ccd-loop1}
Strong couplings of $J/\psi$ to
$D$ and $D^*$.}
\centering\begin{tabular}{|c|cc|cc|}
\hline
Amplitude & $\langle D|\bar c\gamma_\mu c|D\rangle$ & $\langle
J/\psi|\bar c\gamma_\mu\gamma_5d|D\rangle$\ \ &\ $\langle D^*|\bar
c\gamma_\mu c|D\rangle$ & $\langle J/\psi|\bar c\gamma_\mu
d|D\rangle$\\
\hline Form factor &\ $F_+(D\to D)$ & $A_0(D\to\psi)$ &\ $V(D\to D^*)$ & $V(D\to\psi)$\\ 
\hline 
$F(0)$     & $1$             & $0.545\pm 0.003$ & $1.186\pm0.003$ & $1.517\pm0.003$\\ 
$M_R$ & $M_\psi$ & $M_D$ & $M_\psi$ & $M_{D^*}$\\ 
$\sigma_1$ & $0.453\pm0.017$ & $0.58\pm 0.02$   & $0.453\pm0.013$ & $0.59\pm0.01$\\
\hline\ Strong coupling\ \ & \multicolumn{2}{c|}{$g_{DD\psi}=26.04\pm1.43$} &
\multicolumn{2}{c|}{\ $g_{DD^*\psi}=(10.71\pm0.39)\;{\rm GeV}^{-1}$\ \ }\\ \hline\end{tabular}

\caption{\label{table:ccd-loop2}Strong couplings of $\eta_c$ to
$D$ and $D^*$.}
\centering
\begin{tabular}{|c|ccc|c|}\hline\vspace{-2.3ex}&&&
\\
Amplitude & $\langle D|\bar d\gamma_\mu c|\eta_c\rangle$ &
$\langle D^*|\bar d\gamma_\mu\gamma_5c|\eta_c\rangle$ & $\langle
D^*|\bar c\gamma_\mu\gamma_5c|D\rangle$\ \ & $\langle D^*|\bar
d\gamma_\mu c|\eta_c\rangle$ \\
\hline Form factor &\ $F_+(\eta_c\to D)$ & $A_0(\eta_c\to D^*)$ & $A_0(D\to D^*)$ & $V(\eta_c\to D^*)$\\
\hline 
$F(0)$     & $0.643\pm 0.002$ & $0.491\pm0.002$ & $0.966\pm0.004$ & $1.503\pm 0.003$ \\ 
$M_R$ & $M_{D^*}$ & $M_D$ & $M_{\eta_c}$ & $M_{D^*}$ \\ 
$\sigma_1$ & $0.466\pm 0.008$ & $0.71\pm0.01$   & $0.39\pm 0.01$  & $0.491\pm 0.008$ \\
\hline\ 
Strong coupling\ \ & \multicolumn{3}{c|}{$g_{DD^*\eta_c}=15.51\pm0.45$} &\
$g_{D^*D^*\eta_c}=(9.76\pm0.32)\;{\rm GeV}^{-1}$\ \ \\
\hline\end{tabular}\end{table}
We emphasize that the excellent combined description of the sets
of form factors involving the same strong coupling in their pole residues 
(with $\chi^2/{\rm DOF}\le0.1$ assigning a 1\% error
to our form-factor results) lends strong support~to the
reliability of our results. This is actually a highly nontrivial
feature: for instance, the coupling $g_{DD^*\psi}$ is obtained
from a combined description of $V(D\to D^*)$ and $V(D\to \psi)$:
the vector states here have completely different structure and
properties and are described by rather different wave functions.
Also the vector resonances that appear in the~form factors at
time-like momentum transfers differ: $J/\psi$ in $V(D\to D^*)$ and
$D^*$ in $V(D\to \psi)$. The excellent description of all sets of
form factors strongly increases the reliability of our findings. The
behavior of the ``off-shell couplings''~({\em viz.}, the suitably
rescaled form factors equaling the strong
couplings at $q^2=M_R^2$) is depicted in Fig.~\ref{Plot:2}. 

In the limit $m_Q\to\infty$, the form factors $F_+^{D\to D}(q^2)$ and $V^{D\to D^*}(q^2)$ are equal to each other. From
(\ref{residues}), the coupling constants thus satisfy the
heavy-quark symmetry relation $g_{DD\psi}=(M_D+M_D^*)g_{DD^*\psi}$
--- fulfilled with 30\% accuracy.

%**************************************************************************
\section{Strong couplings of $\eta_c$ and $J/\psi$ to $D_s$ and $D_s^*$}
The couplings of $J/\psi $ and $\eta_c$ to the charmed
strange mesons $D_s$ and $D_s^*$ may be found from the residues of
the form factors entering the transition amplitudes induced by the 
currents $\bar c\gamma_\mu c$, $\bar c\gamma_\mu\gamma_5c$, $\bar
d\gamma_\mu c$ or $\bar d\gamma_\mu\gamma_5c$. The
relevant~Feynman diagrams may be inferred from those shown in
Fig.~\ref{Plot:1}a and Fig.~\ref{Plot:1}b by replacing the $d$
quark by the $s$ quark. Tables~\ref{table:ccs-loop1}~and
\ref{table:ccs-loop2} summarize the results of our analysis. Again, we emphasize the excellent simultaneous 
description of the sets of the form factors involving the same strong coupling in the 
pole residues. 

\begin{table}[!hbt]
\caption{\label{table:ccs-loop1}Strong couplings of $J/\psi$ to
$D_s$ and $D_s^*$.}\centering\begin{tabular}{|c|cc|cc|}\hline
Amplitude & $\langle D_s|\bar c\gamma_\mu c|D_s\rangle$ & $\langle
J/\psi|\bar c\gamma_\mu\gamma_5s|D_s\rangle$\ \ &\ $\langle
D_s^*|\bar c\gamma_\mu c|D_s\rangle$ & $\langle J/\psi|\bar
c\gamma_\mu s|D_s\rangle$\\
\hline 
Form factor &\ $F_+(D_s\to D_s)$ & $A_0(D_s\to\psi)$ &\ $V(D_s\to D_s^*)$ & $V(D_s\to\psi)$ \\
\hline $F(0)$ & $1$ & $0.630\pm 0.004$ & $1.23\pm 0.01$ & $1.67\pm 0.01$\\ 
$M_R$ & $M_\psi$ & $M_{D_s^*}$ & $M_\psi$ & $M_{D_s^*}$\\ 
$\sigma_1$ & $0.39\pm 0.01$ & $0.53\pm 0.01$ & $0.39\pm 0.03$ & $0.55\pm 0.02$\\
\hline\ Strong coupling\ \ &
\multicolumn{2}{c|}{\ $g_{D_sD_s\psi}=23.83\pm0.78$\ \ } & \multicolumn{2}{c|}{\ $g_{D_sD_s^*\psi}=(9.60\pm0.80)\;{\rm GeV}^{-1}$\ \ }
\\\hline\end{tabular}

\caption{\label{table:ccs-loop2}
Strong couplings of $\eta_c$ to $D_s$ and $D_s^*$.}
\centering
\begin{tabular}{|c|ccc|c|}\hline
Amplitude & $\langle D_s|\bar s\gamma_\mu c|\eta_c\rangle $ &
$\langle D_s^*|\bar c\gamma_\mu\gamma_5s|\eta_c\rangle$ & $\langle
D_s^*|\bar c\gamma_\mu\gamma_5c|D_s\rangle$\ \ & $\langle
D_s^*|\bar c\gamma_\mu s|\eta_c\rangle$\\
\hline Form factor &\ $F_+(\eta_c\to D_s)$ & $A_0(\eta_c\to D_s^*)$ & $A_0(D_s\to D_s^*)$ & $V(\eta_c\to D_s^*)$ \\
\hline 
$F(0)$ & $0.746\pm 0.002$ & $0.576\pm 0.002$ & $0.953\pm 0.004$ & $1.66\pm 0.004$\\ 
$M_R$ & $M_{D_s^*}$ & $M_{D_s}$ & $M_{\eta_c}$ & $M_{D_s^*}$ \\ 
$\sigma_1$ & $0.42\pm 0.01$ & $0.61\pm 0.01$ &  $0.35\pm 0.01$ & $0.45\pm 0.01$\\
\hline\ Strong coupling\ \ &
\multicolumn{3}{c|}{$g_{D_sD_s^*\eta_c}=14.15\pm0.52$}
&\ $g_{D_s^*D_s^*\eta_c}=(8.27\pm0.37)\;{\rm GeV}^{-1}$\ \ \\
\hline
\end{tabular}
\end{table}

\section{Summary and conclusions}
We revisited the three-meson strong couplings involving $J/\psi$
and $\eta_c$ within the dispersion formulation of the relativistic constituent quark model. 
In this approach, various hadron observables are given by relativistic spectral integrals 
in terms of spectral densities of the relevant Feynman diagrams and of relativistic hadron~wave functions. The
hadron observables from this approach satisfy all rigorous constraints emerging in QCD~in~the heavy-quark limit if
the hadron wave functions are localized in a region of the order of the confinement radius. 
The basic parameters of the model, such as the effective constituent quark masses, 
have been determined before in a study \cite{ms} of heavy-meson transition form factors. 
Following \cite{ms}, we fix the wave-function parameters of $J/\psi$, $\eta_c$, and 
the charmed and charmed strange mesons using the known results for the leptonic decay constants of these
mesons. With these parameters at hand, the form factors of interest are calculated in the space-like region and the weak-decay region 
using relativistic dispersion integrals.    

Our results may be summarized as follows:

\vspace{1ex}\noindent 
1. As the dispersion integrals (\ref{ff}) are based on quark degrees of freedom, all our calculations are carried out far from the 
pole at $q^2=M_R^2$. However, the numerical interpolation formulas turn out to be excellently compatible with the pole at $q^2=M_R^2$ 
and therefore can be used up to $q^2=M_R^2$. This feature allows us to extract the residues of these form factors at $q^2=M_R^2$ 
and to derive in this way the three-meson~couplings. We perform a combined analysis of groups of form factors involving 
the same strong couplings in the pole residues. In all cases~we arrive at an excellent combined description of these form factors
(that is, with $\chi^2/{\rm DOF}\le0.1$, assigning just a 1\% error to our form-factor results). 
This is a highly nontrivial feature, as the same value of the strong coupling is~extracted
from form factors involving mesons which have entirely different wave functions. 
Such excellent description of all~sets of form factors gives strong support to the credibility of our findings.

As summary of our predictions, we~report,
\begin{itemize}
\item for the couplings involving $J/\psi$ and $\eta_c$ mesons,
\begin{align*}
&g_{\eta_c\eta_c\psi}=25.8\pm1.7,\\
&g_{\eta_c\psi\psi}=(10.6\pm1.5)\;{\rm GeV}^{-1},
\end{align*}
\item for the $J/\psi$ and $\eta_c$ couplings to charmed mesons,
\begin{align*}
&g_{DD\psi}=26.04\pm1.43,\\
&g_{DD^*\psi}=(10.7\pm0.4)\;{\rm GeV}^{-1},\\
&g_{DD^*\eta_c}=15.51\pm0.45,\\
&g_{D^*D^*\eta_c}=(9.76\pm0.32)\;{\rm GeV}^{-1},
\end{align*}
\item and, for the $J/\psi$ and $\eta_c$ couplings to charmed
strange mesons,
\begin{align*}
&g_{D_sD_s\psi}=23.83\pm0.78, \\
&g_{D_sD_s^*\psi}=(9.6\pm0.8)\;{\rm GeV}^{-1},\\
&g_{D_sD_s^*\eta_c}=14.15\pm0.52,\\
&g_{D_s^*D_s^*\eta_c}=(8.27\pm0.37)\;{\rm GeV}^{-1}.
\end{align*}
\end{itemize}
The uncertainties quoted in these results are merely the statistical uncertainties related to the accuracy of the 
description of our results by the fit formulas. There are, of course, also systematic uncertainties related to the approximate 
nature of the dispersion approach to the form factors; these uncertainties are very difficult to estimate unambiguously. 
Comparison of the couplings predicted by the dispersion approach \cite{mb,ms} with the results from experiment \cite{cleo,BaBar} 
and lattice QCD \cite{g_lat} in those cases where such results are available, allows us to expect the accuracy 
of our predictions to be not worse than 15--20\%. 

\vspace{1ex}\noindent 2. Our results considerably exceed the ones
from QCD sum rules (see the comparison in
Table~\ref{table:comparison}). Both approaches follow the same
strategy for extracting the strong couplings: the form factors are
calculated in a kinematical region~far away from the pole and are
then extrapolated to the pole region in order to isolate the
residue. The advantage of the dispersion approach for the problem
under consideration is twofold: we predict the form factor in a
broader range~of~$q^2$ and we consider $q^2$ values closer to the
pole region than the region where QCD sum rules may be applied.
Therefore, we need to extrapolate the form factors over much
narrower regions of the momentum transfer and thus believe that
the results of the dispersion approach are more reliable.

\begin{table}[!hbt]
\caption{\label{table:comparison}Comparison of our strong-coupling
predictions with earlier results from QCD sum rules. For
consistency with~definition (\ref{g}), the $PPV$ couplings from
\cite{matheus2005} and \cite{bracco2014} have been multiplied by a
factor 2.}\centering\begin{tabular}{|c|c|c|c|c|}\hline &
$g_{DD\psi}$ &\ $g_{DD^*\psi}$ (GeV$^{-1}$)\ \ & $g_{D_sD_s\psi}$ &\ $g_{D_sD_s^*\psi}$ (GeV$^{-1}$)\ \ \\
\hline 
This work        & $26.04\pm1.43$ & $10.7\pm0.4$ & $23.83\pm 0.78$ & $9.6\pm0.8$
\\\hline\ 
QCD sum rules\ \ &\ $11.6\pm1.8$ \cite{matheus2005}\ \ &
$4.0\pm0.6$ \cite{matheus2005} &\ $11.96\pm1.34$
\cite{bracco2014}\ \ & $4.30\pm1.53$
\cite{bracco2015}\\\hline
\end{tabular}
\end{table}

\vspace{1ex}\noindent 3. We also investigated the $SU(3)$-breaking
effects in the strong couplings. The replacement of the light
quark by the strange quark leads to the increase of the considered
transition form factors and of the corresponding residues. At~the
same time, however, also the leptonic decay constants of the
charmed strange mesons considerably exceed those of their
non-strange counterparts. The three-meson strong couplings are
derived as ratios of the form-factor residues~and the leptonic
decay constants, and, eventually, the replacement of the
non-strange quark by the strange quark leads to a reduction of the
three-meson couplings at the level of some $10\%.$ In contrast,
QCD sum rules observe an enhancement of the three-meson couplings
when the light quark is replaced by the strange one.

\vspace{4ex}\noindent{\bf Acknowledgements.} D.~M.~is grateful to
Berthold Stech for interesting and stimulating discussions.
S.~S.~thanks MIUR (Italy) for partial support under Contract
No.~PRIN 2010--2011.


\begin{thebibliography}{30}
\bibitem{cleo}
S.~Ahmed {\it et al.} (CLEO Collaboration), Phys.~Rev.~Lett.~{\bf 87}, 251801 (2001).
\bibitem{lin2000}
Z.~Lin and C.~M.~Ko, Phys.~Rev.~C {\bf 62}, 034903 (2000).
\bibitem{bracco2005}
M.~E.~Bracco, M.~Chiapparini, F.~S.~Navarra, and M.~Nielsen, Phys.~Lett.~B {\bf 605}, 326 (2005).
\bibitem{matheus2005}
R.~Mattheus {\it et al.}, Int.~J.~Mod.~Phys.~E {\bf 14}, 555 (2005).
\bibitem{bracco2014}
B.~Osorio Rodriguez, M.~E.~Bracco, and
M.~Chiapparini, Nucl.~Phys.~A {\bf 929}, 143 (2014).
\bibitem{bracco2015}B.~Osorio Rodriguez {\it et al.},
Eur.~Phys.~J.~A {\bf 51}, 28 (2015).
\bibitem{braun}V.~M.~Belyaev, V.~M.~Braun, A.~Khodjamirian, and
R.~R\"uckl, Phys.~Rev.~D {\bf 51}, 6177 (1995).
\bibitem{da}
D.~Melikhov, Eur.~Phys.~J.~direct {\bf C4}, 2 (2002) [arXiv:hep-ph/0110087].
\bibitem{mb}
D.~Melikhov and M.~Beyer, Phys.~Lett.~B {\bf 452}, 121 (1999).
\bibitem{ms}
D.~Melikhov and B.~Stech, Phys.~Rev.~D {\bf 62}, 014006 (2000).
\bibitem{cqm}
A.~Le Yaouanc, L.~Oliver, S.~Ono, O.~Pene, and J.-C.~Raynal, 
Phys.~Rev.~D {\bf 31}, 137 (1985); Phys.~Rev.~Lett.~{\bf 54}, 506 (1985); 
W.~Lucha, F.~Sch\"oberl, and D.~Gromes, Phys.~Rep.~{\bf 200}, 127 (1991); 
F.~Cardarelli, E.~Pace, G.~Salm\`e, and S.~Simula, Phys.~Lett.~B {\bf 357}, 267 (1995); 
R.~N.~Faustov and V.~O.~Galkin, Z.~Phys.~C {\bf 66}, 119 (1995); 
D.~Ebert, R.~N.~Faustov, and V.~O.~Galkin, Phys.~Lett.~B {\bf 635}, 93 (2006).
\bibitem{gv}
D.~Melikhov and B.~Stech, Phys.~Rev.~D {\bf 74}, 034022
(2006); W.~Lucha, D.~Melikhov, and S.~Simula, Phys.~Rev.~D {\bf
74}, 054004 (2006).
\bibitem{anis}
V.~V.~Anisovich, D.~I.~Melikhov, and V.~A.~Nikonov,
Phys.~Rev.~D {\bf 55}, 2918 (1997); 
A.~F.~Krutov and V.~E.~Troitsky,~JHEP {\bf 9910}, 028 (1999).
\bibitem{melikhov}
D.~Melikhov, Phys.~Rev.~D {\bf 53}, 2460 (1996); {\bf 56}, 7089
(1997).
\bibitem{constituent}
M.~S.~Bhagwat, M.~A.~Pichowsky, C.~D.~Roberts, and P.~C.~Tandy, Phys.~Rev.~C {\bf 68}, 015203 (2003); 
G.~Eichmann, R.~Alkofer, I.~C.~Cl\"oet, A.~Krassnigg, and C.~D.~Roberts, Phys.~Rev.~C {\bf 77}, 042202 (2008); 
D.~Melikhov and S.~Simula, Eur.~Phys.~J.~C {\bf 37}, 437 (2004).
\bibitem{pdg}
K.~A.~Olive {\it et al.} (Particle Data Group),
Chin.~Phys.~C {\bf 38}, 090001 (2014).
\bibitem{lmsfD}
W.~Lucha, D.~Melikhov, and S.~Simula, Phys.~Lett.~B
{\bf 701}, 82 (2011).
\bibitem{lmsfD*}
W.~Lucha, D.~Melikhov, and S.~Simula, Phys.~Lett.~B
{\bf 735}, 12 (2014).
\bibitem{fDDs*lat}
D.~Be\v{c}irevi\'c {\it et al.}, JHEP {\bf 1202}, 042 (2012).
\bibitem{fetac_lat}
C.~T.~H.~Davies, C.~McNeile, E.~Follana, G.~P.~Lepage, H.~Na, and J.~Shigemitsu, 
Phys.~Rev.~D {\bf 82}, 114504 (2010).
\bibitem{fpsi_lat}
G.~C.~Donald, C.~T.~H.~Davies, R.~J.~Dowdall, E.~Follana, K.~Hornbostel, J.~Koponen, 
G.~P.~Lepage, C.~McNeile, Phys.~Rev.~D {\bf 86}, 094501 (2012).
\bibitem{V(0)exp}
R.~E.~Mitchell {\it et al.} (CLEO Collaboration), Phys.~Rev.~Lett.~{\bf 102}, 011801 (2009).
\bibitem{V(0)lat}
D.~Be\v{c}irevi\'c and F.~Sanfilippo, JHEP {\bf 1301}, 028 (2013).
\bibitem{BaBar}
J.~P.~Lees {\it et al.} (BaBar Collaboration), Phys.~Rev.~Lett. {\bf 111}, 111801 (2013). 
\bibitem{g_lat}
D.~Be\v{c}irevi\'c and F.~Sanfilippo, Phys.~Lett.~B {\bf 721}, 94 (2013); 
J.~M.~Flynn {\it et al.}, arXiv:1506.06413 [hep-lat].
\end{thebibliography}
\end{document}